# Large diameter TiO$_2$ nanotubes enable integration of conformed hierarchical and blocking layers for enhanced dye-sensitized solar cell efficiency

Abdelhamid Elzarka, [a, b] Ning Liu,[a] Imgon Hwang, [a] Mustafa Kamal,[b] and Patrik Schmuki*[a, c]

[a] A. Elzarka, Dr. N. Liu, I. Hwang, Prof.Dr. P. Schmuki Department of Materials Science WW-4, LKO University of Erlangen-Nuremberg Martensstrasse 7, 91058 Erlangen (Germany) Fax: (+49) 9131-852-7582 E-mail: schmuki@ww.uni-erlangen.de

[b] A. Elzarka, Prof.Dr. M. Kamal Metal Physics Laboratory, Physics Department Faculty of Science, Mansoura University, Mansoura 35516 (Egypt)

[c] Prof.Dr. P. Schmuki Department of Chemistry, King Abdulaziz University, Jeddah (Saudi Arabia)

**Abstract**

In the present work we grow anodic TiO$_2$ nanotube layer with tube diameter ~ 500 nm and an open tube mouth. We use this morphology in dye-sensitized solar cells (DSSCs) and show that these tubes allow the construction of hybrid hierarchical photoanode structures of nanotubes with a defined and wall-conformance TiO$_2$ nanoparticles decoration. At the same time, the large diameter allows the successful establishment of an additional (insulating) blocking layer of SiO$_2$ or Al$_2$O$_3$. We show that this combination of hierarchical structure and blocking layer significantly enhances the solar cell efficiency by suppressing recombination reactions. In such a DSSC structure, the solar cell efficiency under back side illumination with AM1.5 illumination is enhanced from 5% neat tube to 7 %.

**Introduction**

The first prototype of dye-sensitized solar cells (DSSCs) introduced by O'Regan and Grätzel in 1991 have attracted a great of interest for more than two decades owing to the potential low cost alternative to traditional silicon solar cells [1]. DSSCs mainly consist of a nanocrystalline photoanode film consisting of TiO2 anatase particles covered by a monolayer of dye molecules, a redox electrolyte, and a counter electrode. Considerable efforts have been devoted to the development of more efficient photoanode materials by combining the merits of the large surface area of nanoparticles (NP) with the straight transport path of one-dimension geometries (nanorods, nanowires, nanotubes) [2]–[14].

The main path of losses in DSSCs is the recombination between photogenerated electrons with the oxidized dye and the redox species in the electrolyte [15][16]. One way for reducing this recombination pathway is to apply onto the TiO2 electron transport material a very thin layer of insulator or semiconductor with a higher band gap, specifically with a conduction band edge higher in energy than TiO2 which acts as a so-called electron blocking layer.

In classic nanoparticle based photoanodes of TiO2, large range of semiconductors have been investigated as a thin blocking layer coating, such as SrCO3 [17], Al2O3 [18]–[23], SrTiO3 [24][25], SiO2 [20], [26], Ga2O3 [27], Nb2O5 [28], MgO [23][29], ZrO2 [30][31] and ZnO[23]. Typically such blocking layers were found to enhance the overall DSSC efficiency either by enhancing the short circuit current and/or decreasing the open circuit potential – prominent examples the use of thin Al2O3 or SiO2 layers encapsulate TiO2 nanoparticles with a tunneling barrier[20].

While the introduction of such blocking concepts has been used for NP-based SS it has not been explored for TiO2 NT assemblies. One of the key challenges is to be able to "engineer" the inner walls of the tubes with sufficient control. While anodic NTs grown on their metallic Ti substrates represent an ideal electrode geometry with a desired directionality, usually their specific surface area is clearly lower than conventional nanoparticle photoanodes. This drawback can basically be overcome by establishing TiO2 nanoparticles decorated tube-walls (hierarchical structure) and as outlined above on additional thin layer coating with a blocking layer would be desired. However, in many cases self-organized TiO2 NT arrays carry either "initiation layers", "grassy-morphologies" or show a narrowed mouth [20, 21]. All these geometries significantly affect a proper filling or defined decoration of the tube walls. Ideally tubes should have a large diameter and have entirely open tube mouth to allow a fully controlled tube wall modification.

In the present work we introduce a morphology of titania nanotube array with a diameter of ~ 500 nm and length 12 μm that provide an open tube mouth. These tubes can be layer-by-layer decorated with TiO2 NPs forming a hierarchical high surface structure, in particular, benefit it allows to additionally establish a thin blocking layers of SiO2 or Al2O3 that significantly affect the DSSC performance.

**Results and discussion**

After a set of exploratory experiments as established an anodization treatment as described in the experimental section to yield to self-organized nanotube layer with a desired large diameter. These tubes as shown in the SEM images of Fig. 1. Have a diameter ~ 500 nm and a tube length of 12 µm. This large diameter is ascribed to the application of a high voltage (170 V) without triggering break down events. Important is that these tubes provide an opened tube top without initiation layer or 'grass' morphology which is due to the mild etching rate of the tube to particles in the presence of the comparable high concentration of F$^-$ ions. These tubes were then annealed in air at 450 °C in order to convert the crystallized material as anatase. In order to fabricate hierarchical structure usning a TiCl4 treatment as describes in the SI.

For higher DSSCs efficiency by TiCl4 treatments, we applied different thickness of TiO2 nanoparticle layers by repeated dipping the nanotubes array for different times and different concentrations of TiCl4 solution. The particle size of the TiO2 nanoparticles was found to depend on the concentration of TiCl4 in solution. To obtain smaller particles (yielding a higher surface area) we found that dipping NT 6-times in 0.2M TiCl4 aqueous solution followed by 3-times in 0.1M TiCl4 aqueous solution is an optimized condition for a highest DSSCs efficiency by TiCl4 treatments, see fig. (1, B). Under these condition for decoration, the nanotube wall is decorated on both sides (inside and outside surfaces) and the wall thickness increased from 20 nm to approx. 230 nm while the tube top still remaining opened.

To create a thin blocking layer of SiO2 or Al2O3 the annealed samples (450 °C for 20 min) were dipped under various conditions in tetraethylorthosilicate or Aluminum-tri-sec-butoxide in dry isopropanol, respectively. The optimized treatment was found for SiO2 in 75 mM of tetraethylorthosilicate in dry ethanol for 1h at 70 °C and Al2O3 in 75 mM of Aluminum-tri-sec-butoxide in dry isopropanol for 10min at 70 °C. Then all the samples are rinsed with ethanol, dried with nitrogen stream and finally are sintered at 450°C for 10 min.

EDX, XPS and TOF-SIMS were performed for chemical composition characterization. As shown in Fig. (2) and table (1), EDX analysis shows the presence of Ti, O and C in all samples and yields for Si and Al on atomic concentration of 0.56% and 1.31% for the NT+NP+SiO2 and NT+NP+Al2O3 samples respectively.

XPS peaks shown in Fig. (2) exhibit the signals for Si and Al. The peak of Si2p is located at 102.5 eV which confirms the presence of $SiO_2$ ($Si^{4+}$) [32]–[34]. The evaluation of the concentration yields 0.5 %, which is in good agreement with SEDX. I.e. the concentration provided from the tube tops (XPS) is in line with the bulk value of EDX – thus indicating a uniform coating of the tubes in the depth of the tubes. In the case of alumina XPS indicates same on richment of the mouth but …The Al2p peak is located at 74.3 eV which confirms doping the surface with $Al^{3+}$ and forming a thin oxide layer of $Al_2O_3$[35]. In accord with the presence of O 1s peaks shows a small shift of $SiO_2$ and $Al_2O_3$ respectively to lower binding energy for NT+NP+$SiO_2$ and NT+NP+$Al_2O_3$ samples.

Fig. (2) shows TOF-SIMS measurement which confirms also the presence of Si and Al on the surface of the samples after the treatment with Si and Al precursors. we can also notice that the intensity of Ti and TiO decrease after treatment with Si and Al precursors, which ensure the presence of $SiO_2$ or $Al_2O_3$ coating layer at the expense $TiO_2$.

Fig. (3) shows solar cell data taken under back side illumination conditions and Table 2 gives the extracted solar cell performances. The overall DSSC efficiency η by applying $SiO_2$ insulating layer shows enhancement about 60% and 15% for samples treated without and with $TiO_2$ nanoparticles respectively. The short circuit current $I_{sc}$ for samples NT and NT+NP treated with $SiO_2$ has been enhanced by > 10%. Also for nanotubes SiO2 treated without TiO2 nanoparticles, the open circuit potential Voc is increased by nearly 10%.

In the case of Al₂O₃ coating, for NT+Al₂O₃, there is nearly 10%, 4% and 15% increase in $V_{oc}$, $I_{sc}$ and η respectively. Also for NT+NP+Al₂O₃, $I_{sc}$ and η increase by 11% and 9% than NT+NP respectively, which show nearly the same trend like the results published by Alarcón et al. [21] and Kim et al. [22] for nanoparticle layer, see Fig. (3) and Table 2.

As shown in Table 2, we can see the significant increase in dye loading after decoration with TiO₂ nanoparticles. Although SiO₂ is more acidic and Al₂O₃ is more basic than TiO₂ (pH=2.1, 9.2 and 5.5 respectively [20]), the dye loading for samples treated without/with SiO₂ or Al₂O₃ are nearly the same. This means that there is no change in the light harvesting efficiency by adding a thin layer of SiO₂ or Al₂O₃. This confirms the deduction that the recombination rate is decreased to the lowest level in case of NT+NP+SiO₂ and NT+NP+ Al₂O₃.

In order to study in detail the recombination and transport properties of the samples and explain the efficiency results, intensity-modulated photocurrent spectroscopy (IMPS) and intensity-modulated photovoltage spectroscopy (IMVS) measurements have been investigated. Fig. (4) shows the electron transport time constant $\tau_c$ and the recombination time constant $\tau_r$ as a function of various light intensities. The results show that both $\tau_c$ and the $\tau_r$ decrease along with the increasing light intensities, which could be ascribed to the influence of the trapping/detrapping processes under high light illumination condition[38]. The electron transport time constant $\tau_c$ is lower for samples treated with with SiO₂ or Al₂O₃ insulating layer ( NT+NP+SiO₂ and NT+NP+Al₂O₃) which means that electrons transport is faster with SiO₂ or Al₂O₃ coating (Fig. 4, A). Also the recombination time constant $\tau_r$ is higher for samples coated with SiO₂ or Al₂O₃. This means that the electron probability for surviving from recombination is higher with SiO₂ or Al₂O₃ coating (Fig. 4, B).

Electron diffusion coefficient $D_n$ extracted from electron transport time constant $\tau_c$ is higher for samples treated with SiO₂ or Al₂O₃ coating, see Fig. (4). So that photogenerated electrons can diffuse for longer path inside the photoanode and increase their probability to reach the back contact without facing recombination.

OCVD measurements [Fig. (12, A)] show that the samples treated with bloking layer of $SiO_2$ or $Al_2O_3$ have the lowest OCVD or in other words have the lowest electron-hole recombination rate. This indicates that the electrons injected from excited dye can survive longer (e.g. longer lifetime) and hence can facilitate electron transport without undergoing losses by recombination with the redox species inside the electrolyte, see Fig. (12, B).

The present results show that both $Al2O3$ and $SiO2$ can have a benefit effect on TiO2 nanotubes solar cells. This may be ascribed to the blocking layer introduced at the interface between the back contact substrate and the electrolyte. Since the nanotube layer might be detached from the substrate at some places and give the chance for redox species presented in the electrolyte to recombine with electrons collected at the back contact substrate, $SiO_2$ or $Al2O3$ prevent recombination between redox species and electrons collected at the back contact substrate.

This large effect when thin layer of $SiO_2$ is applied directly on NT samples may be related not only to plain layer properties of $SiO_2$ (insulating blocking layer), but also the modification of the junction between $SiO_2$ and $TiO_2$. As for example, Müller et al. [36] revealed that the formation of Ti-O-Si linkages, anodic shift of band edges, and a broadening of $TiO_2$ band gap.

Since the trivalent state $Al^{3+}$ is the most common ionized state for aluminum, so the surface treatment has been reported to form with Al precursor leads to doping the surface and replacing $Ti^{4+}$ with $Al^{3+}$ besides forming thin oxide layer of $Al_2O_3$ on the surface. Replacing $Ti^{4+}$ with $Al^{3+}$ means reducing the oxygen vacancies near the surface which is considered as one of the main factors causing charge recombination in DSSC.

O'Regan et al. introduced three ways that such surface coating can cause an increase in the Voc of a dye-sensitized cell. One way of them said that "the insulating nature of the coating material means that photoinjected electrons in the $TiO_2$ can only recombine with a positive charge in the electrolyte or hole conductor by tunneling through the insulator. The barrier causes a decrease in the "per electron" recombination rate constant for a given electron population. If, at 1 sun illumination, the flux of injected electrons from the dye is unchanged, then the electron concentration at $V_{oc}$ will be higher for the cell with the coating. A larger electron concentration in the $TiO_2$ means a more negative Fermi level and thus a larger Voc"[37].

We claim that a thin layer of insulating semiconductors like $SiO_2$ or $Al_2O_3$ can reduce the electrostatic attraction force between negative photogenerated electrons and oxidized species in the electrolyte and acts as a shielding layer between them.

**Conclusions**

In this work we produced for the first time wide $TiO_2$ nanotube diameter (on average 550 nm) with length around 12 μm. we used this new morphology in DSSC and decorated it with thick layer of $TiO_2$ nanoparticles by hydrolysis method of $TiCl_4$ to increase the surface area. This increase in surface area of the photanode leads to higher dye loading which cause a big enhancement in DSSC efficiency. Further treatment has been done by depositing a thin insulating layer $SiO_2$ or $Al_2O_3$ on the surface of the photoanode. EDX, XPS and TOF-SIMS confirm the presence of Si and Al on the surface of the treated samples. This insulating layer reduce the recombination rate which takes place between photogenerated electrons and oxidized species in the electrolyte and leads to enhancement in η by 15% for NT+NP+$SiO_2$ and 9% for NT+NP+$Al_2O_3$. After these treatment the DSSC efficiency reaches ~ 7% under 1 sun (AM 1.5) back side illumination.

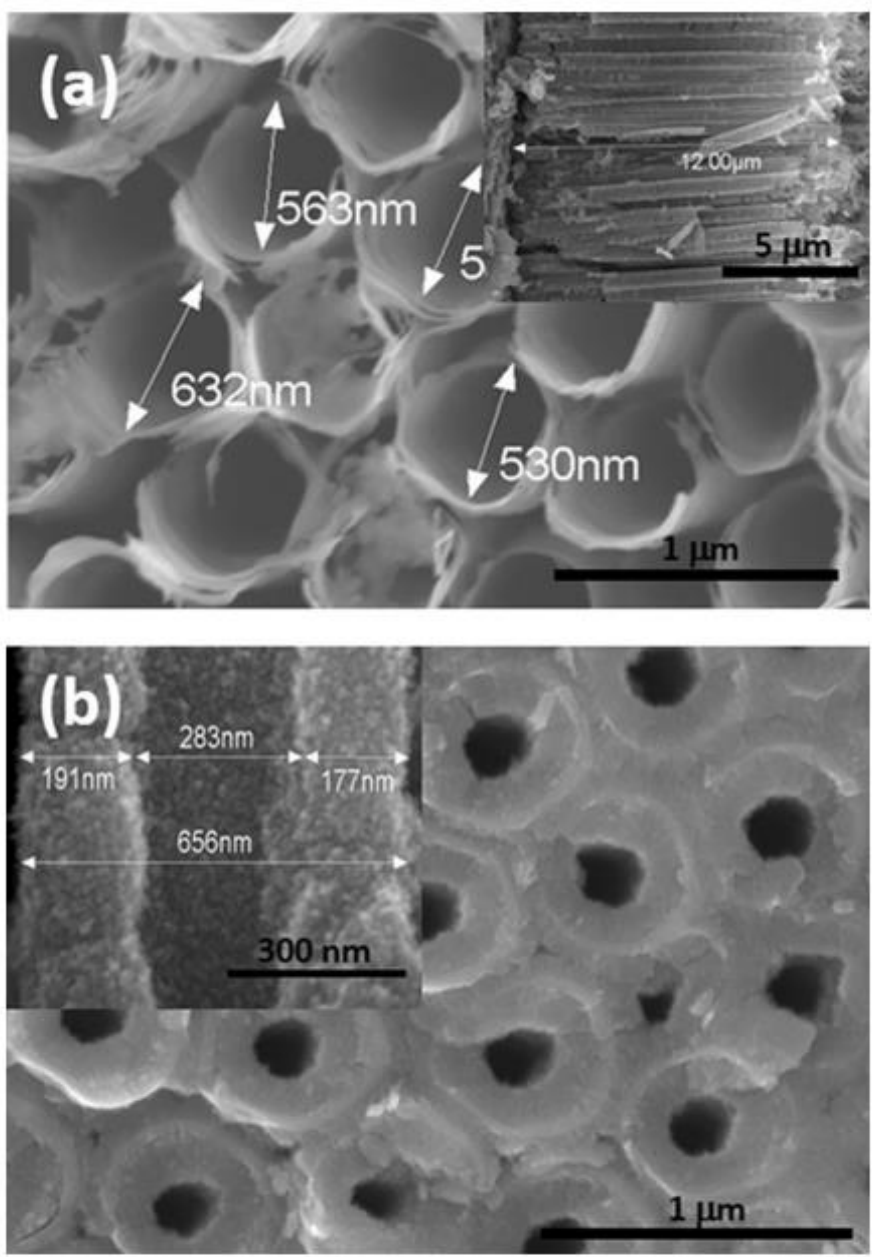

**Figure 1.** (a) Top view of TiO$_2$ nanotube anodized in 85 vol% EG-15 vol% H$_2$O 0.3 M NH$_4$F at 170 V for 4 h (inset: the cross-section of the nanotubes). (b) Top view of tubes after treatment with TiO$_2$ nanoparticles by dipping six times with 0.2 M TiCl$_4$ solution followed by three times with 0.1 M TiCl$_4$ solution (inset: the cross-section of one tube after decoration).

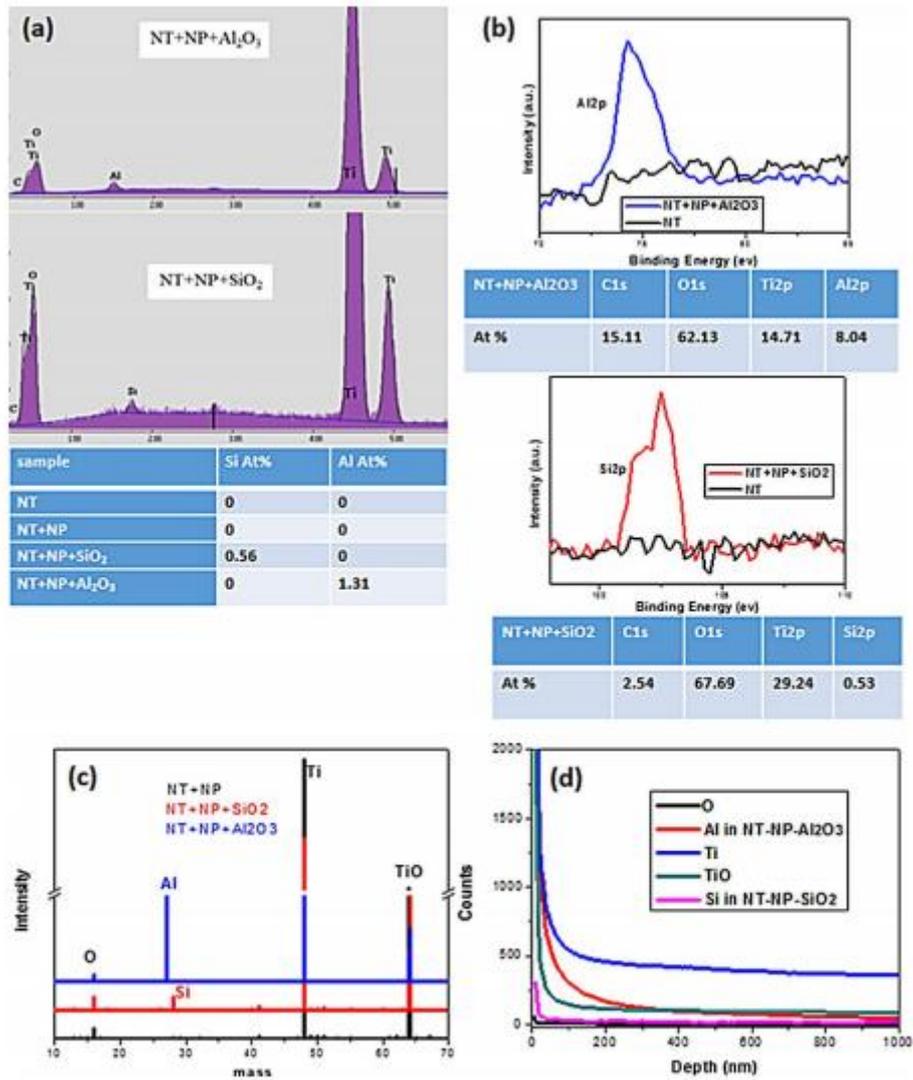

**Figure 2.** (a) EDX for NT + NP + SiO$_2$ and NT + NP + Al$_2$O$_3$ (Table: composition of Si and Al content extracted from EDX results). (b) XPS peaks of Si2p and Al2p for samples treated with SiO$_2$ or Al$_2$O$_3$ insulating layer (Table: composition of Si and Al content extracted from XPS results). (c) and (d) TOF-SIMS and depth profile for NT + NP, NT + NP + SiO$_2$ and NT + NP + Al$_2$O$_3$.

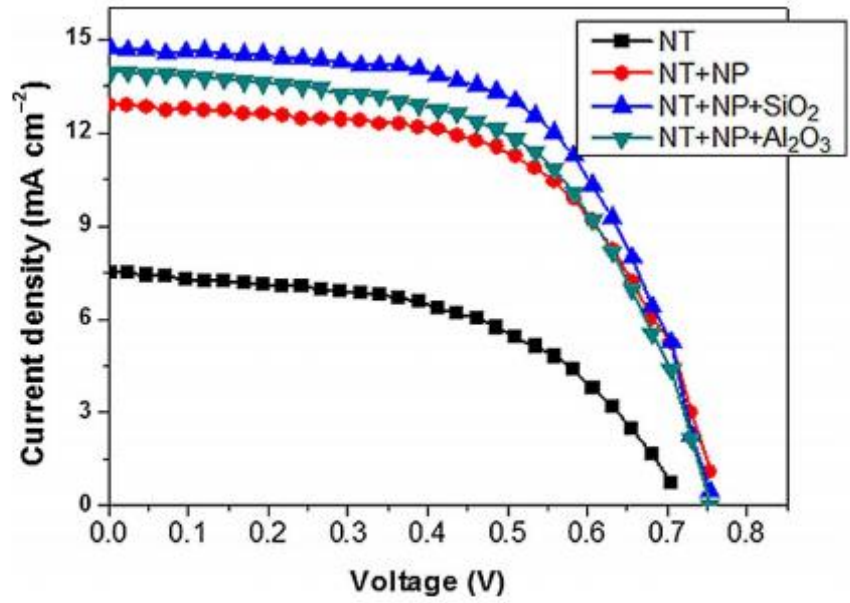

**Figure 3.** *I/V* characteristics of dye-sensitized solar cells under back-side illumination with simulated AM1.5 light.

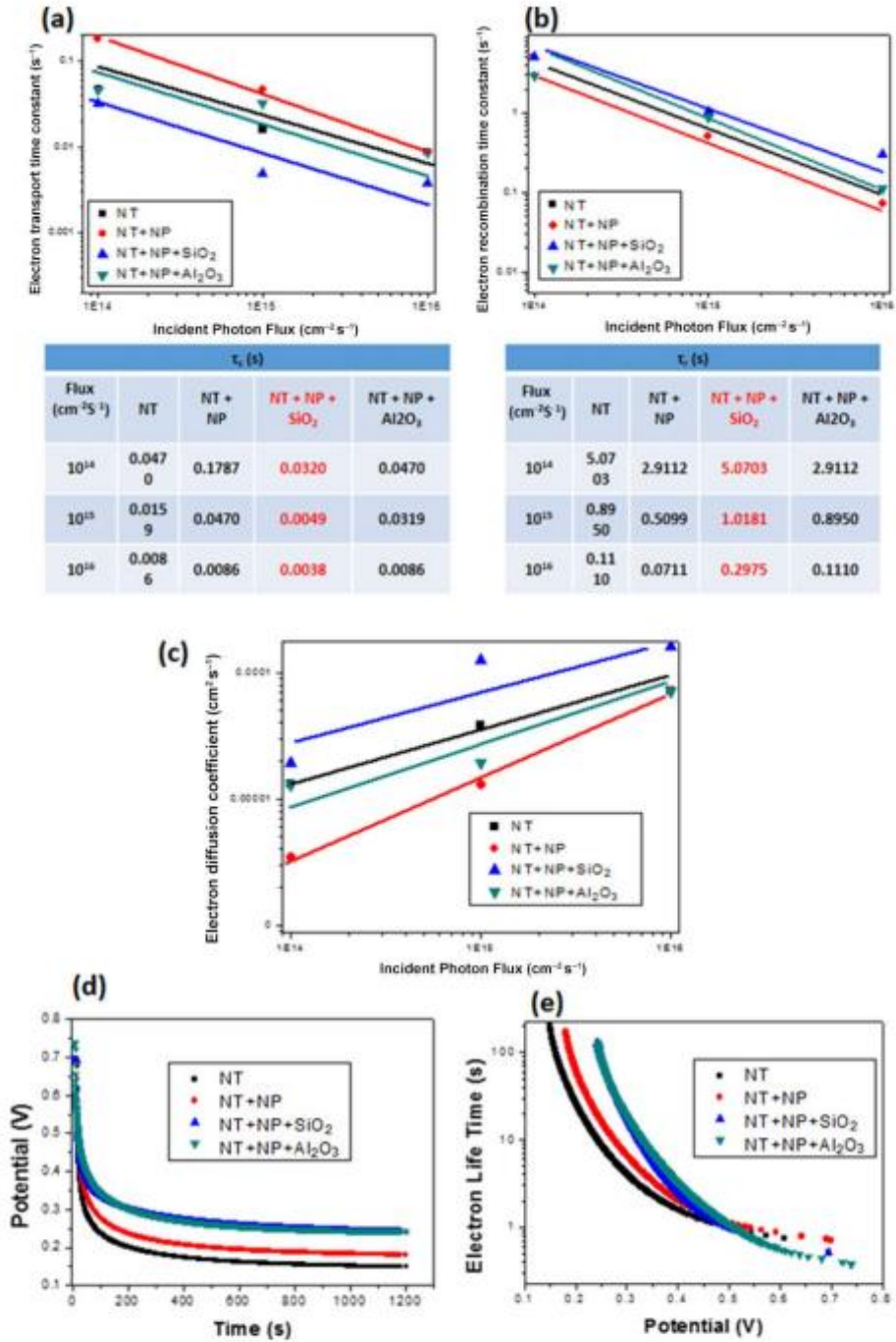

**Figure 4.** (a) Transport-time constant ($\tau_c$). (b) Recombination-time constant ($\tau_r$) as a function of the incident photon flux under 530 nm laser extracted from fitted optical impedance data (IMPS) and (IMVS), respectively. (c) Electron diffusion coefficient $D_n$ as a function of the photon flux density. (d) Photovoltage decay measurements of solar cells. (e) Electron life time curves as a function of open circuit potential.